\documentclass[reprint,aps,showpacs,floatfix,notitlepage,superscriptaddress,amsmath,amssymb]{revtex4-2}

\usepackage{float}
\usepackage{graphicx}
\usepackage{rotating}
\usepackage[toc,page]{appendix}
\usepackage{dcolumn}
\usepackage{bm}
\usepackage{soul}
\usepackage[colorlinks=true,allcolors=blue]{hyperref}
\usepackage{color}
\usepackage{xcolor}
\newcommand{\p}[1]{\textbf{#1}}
\newcommand{\bnabla}{\bm{\nabla}}

\newcommand{{\pell}}{\boldsymbol{\ell}}
\newcommand{\ee}{\textrm{e}}
\newcommand{\cc}{\textrm{c}}
\newcommand{\cP}{{\cal P}}
\let\epsilon\varepsilon
\usepackage{bm}

\newcommand{\B}{\mathrm{B}}
\let\epsilon\varepsilon
\usepackage[normalem]{ulem}

\begin{document}

\title{Mechano-chemical control of enzymatic activity in crowded environments}

\author{Michalis Chatzittofi}
\email{mc2623@cam.ac.uk}
\affiliation{Max Planck Institute for Dynamics and Self-Organization (MPI-DS), D-37077 G\"ttingen, Germany}
\affiliation{DAMTP, Centre for Mathematical Sciences, University of Cambridge, Cambridge CB3 0WA, United Kingdom}

\author{Luca Cocconi}
\affiliation{Max Planck Institute for Dynamics and Self-Organization (MPI-DS), D-37077 G\"ttingen, Germany}
\affiliation{TCM Group, Cavendish Laboratory, University of Cambridge, Cambridge CB3 0US, United Kingdom}

\author{Jakob Metson}
\affiliation{Max Planck Institute for Dynamics and Self-Organization (MPI-DS), D-37077 G\"ttingen, Germany}

\author{Ramin Golestanian}
\email{ramin.golestanian@ds.mpg.de}
\affiliation{Max Planck Institute for Dynamics and Self-Organization (MPI-DS), D-37077 G\"ttingen, Germany}
\affiliation{Rudolf Peierls Centre for Theoretical Physics, University of Oxford, Oxford OX1 3PU, UK}

\date{\today}

\begin{abstract}
Macromolecular crowding has long been known to alter biochemical equilibria
and enzyme kinetics, with reported catalytic activities that may increase, decrease, or vary non-monotonically with crowder concentration. Classical Michaelis--Menten kinetics does not by itself
identify which of these responses should occur for a conformationally active enzyme. Here, we develop a minimal non-equilibrium mechano-chemical theory for a fueled enzyme embedded in a bath of passive crowders and uncover a fluctuation-induced effective force dipole acting on the enzyme subunits, akin to a generalized form of dispersion or depletion interaction. Since the preferred conformation of the model enzyme varies through the catalytic cycle, the fluctuation-induced mechanical load affects the relevant kinetic activation barriers
and is sufficient to produce both monotonic inhibition and non-monotonic modification of turnover.
The framework provides a thermodynamically consistent route for linking crowded environments to catalytic currents, suggesting how mechanical design principles could be leveraged to control enzymatic pathways.
\end{abstract}
\maketitle
The cytoplasm is a heterogeneous, volume-occupied medium containing proteins, enzymes, nucleic acids, metabolites and other macromolecular assemblies~\cite{Ellis2001,Rivas2016,ZhouRivasMinton2008}. This ``macromolecular crowding'' affects biochemical processes through excluded volume, nonspecific attractive and repulsive interactions, hydrodynamic interactions, confinement, and changes in transport~\cite{Minton1981Oct,HuJiangRajagopalan2007,Minton2013StericChemical,AndoSkolnick2010}. These physical effects coexist with signaling, metabolic feedback, and the self-organization of biomolecular condensates~\cite{Wang2011May,Hyman2014,Banani2017May,Speer2021Mar,Testa2021,Dindo2024Jun,Losa2022Apr,Ouazan-Reboul2023Jul,Kokkoorakunnel2024}. A central challenge is therefore to identify the key mechanisms by which the mechanical state of a crowded environment can be coupled to chemical reaction currents.

Enzyme-crowding experiments show that the effect of the crowding environment on the enzyme activity can be complex. Earlier surveys report enzyme activities that increase, decrease, remain almost unchanged, or increase and then decrease as crowder concentration is varied~\cite{ZhouRivasMinton2008}. More recent enzyme-specific studies reach the same conclusion: crowding can slow alcohol dehydrogenase kinetics through diffusion and excluded-volume effects~\cite{Schneider2015ADHCrowding}, decrease citrate synthase activity in a way that depends on polymer overlap and crowder identity~\cite{Wilcox2020CitrateSynthase}, or enhance adenylate kinase activity over part of the concentration range while reducing it at higher concentration~\cite{RastogiChowdhury2021}. Similarly, experiments in depletion-induced enzymatic condensates reveal that lactate dehydrogenase activity varies non-monotonically with the concentration of bovine serum albumin protein crowders inside a droplet~\cite{Dindo2025Jan}. Motivated by these observations, we ask the following mechanistic question: under what conditions can the same crowding load enhance one enzyme and inhibit another, or enhance and inhibit different parts of the same catalytic cycle? We address this question via a systematic theoretical investigation, using a minimal model in which the mechanical and chemical degrees of freedom are coupled through a thermodynamically consistent framework.

Enzymes are nano-scale machines capable of accelerating reactions and using free energy to drive thermodynamically unfavorable processes~\cite{Borsley2024Jun}. The classical barrier crossing picture of Kramers~\cite{kramers1940brownian}, together with Michaelis-Menten kinetics~\cite{MichaelisMenten,johnson2011original}, remains a powerful minimal language for catalytic rates. Yet single-molecule, structural, and computational studies show that enzymatic turnover is frequently accompanied by conformational fluctuations and domain motions~\cite{English2005,Sherman2008,glowacki2012taking,Yeh2023}. Crowding can reshape these conformational landscapes by favoring compact states, changing open--closed population ratios, modifying potentials of the mean force, and altering transition-state barriers~\cite{DongQinZhou2010,Yuan2008MechanicalStability,CandottiOrozco2016}. Similar effects are also present in other nano-scale machines such as ion channels \cite{Bonthuis2014}.

These observations suggest a natural non-equilibrium mechano-chemical viewpoint. The enzyme transduces chemical energy into mechanical displacements, much like a micro- or nano-motor~\cite{RevModPhys.69.1269,PhysRevLett.72.2656,Ehrmann2025Jun,Chatzittofi2025May}. Its functionality may then be described by at least two coupled degrees of freedom: a chemical reaction coordinate driven by a non-equilibrium washboard potential and a spatial conformational coordinate coupled to it~\cite{jaime}. Such models have explained aspects of enzyme enhanced diffusion~\cite{Muddana2010Feb,Sengupta2013Jan,AgudoCanalejo2018,Illien2017} and, when thermodynamic constraints are enforced, lead to mechanistic rules for enzyme design~\cite{Chatzittofi2025May}. They also go beyond a purely quasi-equilibrium insertion or transfer-free-energy picture by retaining an explicitly dissipative reaction cycle.

In this Letter, we develop a minimal theoretical framework that isolates the effect of a crowding-induced mechanical load on a conformationally active, fueled enzyme. Our theory respects thermodynamically consistent enzyme kinetics~\cite{Wachtel2018Apr,RevModPhys.97.015002} and uses a continuous cyclic reaction coordinate driven through a non-equilibrium washboard potential~\cite{jaime,Chatzittofi2025,PhysRevE.100.012404}. We incorporate the enzymatic mechanical degrees of freedom and chemically-induced conformational changes, and perform a systematic coarse-graining over the crowder degrees of freedom while taking into account the enzyme-crowder mechanical and hydrodynamic interactions. The coarse-graining allows us to derive the effective theory for the enzymatic degrees of freedom, in which we find a force-dipole contribution that is proportional to the average density of crowder proteins. This fluctuation-induced interaction is analogous to dispersion or depletion interactions \cite{Kardar1999}. The mechanical coupling to the environment renormalizes activation barriers around the cycle, and thereby produces monotonic inhibition or non-monotonic turnover, depending on the phase relation between the enzyme conformational changes and the chemical barriers.

\begin{figure*}[t]
\centering\includegraphics[width=0.75\linewidth]{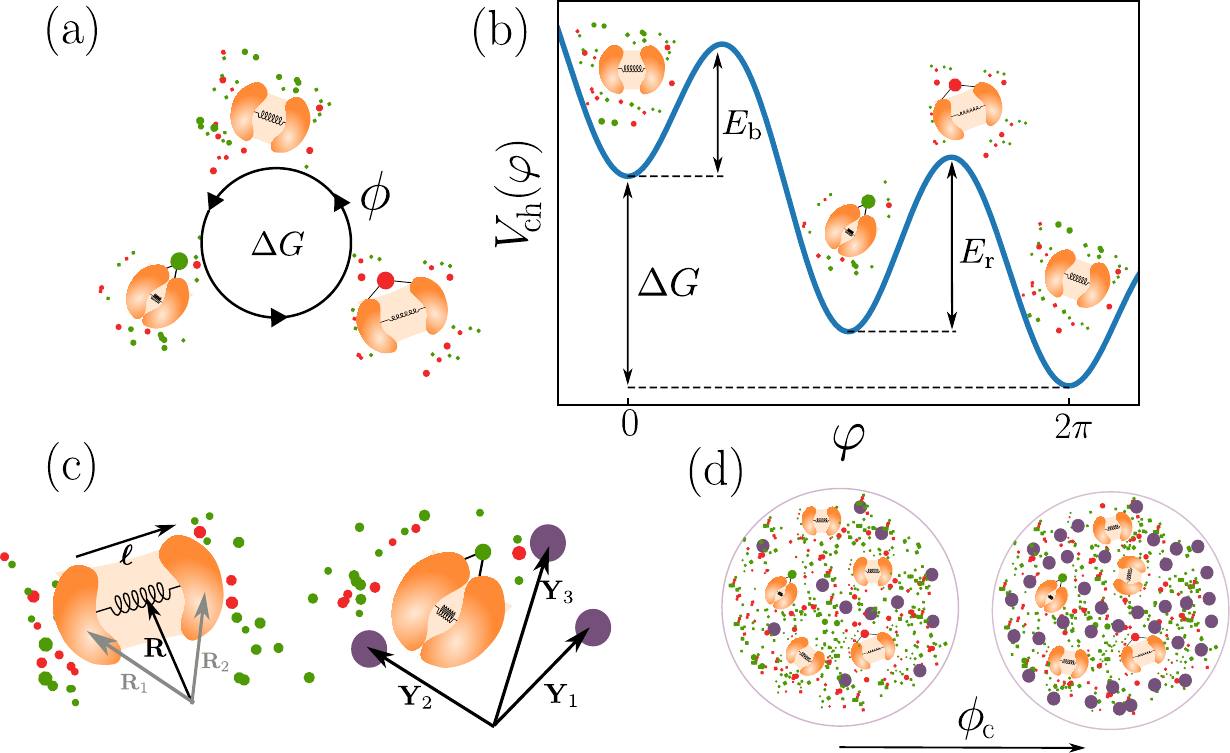}
\caption{Coupling chemical and mechanical coordinate dynamics for a single enzyme in crowded and complex environments. (a) A non-equilibrium washboard potential $V_\mathrm{ch}(\varphi)$, whose tilt is associated to the Gibbs free-energy $\Delta G$ released after a reaction. The different energy barriers indicate the existence of different stages during the whole enzymatic cycle. (b) The enzymatic cycle described by a reaction coordinate driven by $\Delta G$. The three classical stages of an enzymatic reaction; first binding a substrate (green dot) that is converted into a product (red dot) and is eventually released.
(c) Coordinate systems of the enzyme sub-units and the position coordinates of the proteins (purple spheres).
(d) Examples of condensates with different concentrations $\phi_\cc$ of crowding proteins. }
\label{fig:fig1}
\end{figure*}

We begin by considering a dumbbell-like enzyme consisting of two connected non-identical subunits, which serves as a coarse-grained description where an enzyme's complex structure~\cite{AdelekeLarodo2019} is reduced to a single conformational degree of freedom representing the softest relevant principal component mode of the enzyme conformations. This allows us to minimally take into account the resulting internal forces, and also the fluctuating conformational dynamics of an enzyme.
To model the catalytic cycle of the enzyme (for instance a substrate-to-product reaction), we use a periodic coordinate $\varphi$ governed by a washboard potential $V_\mathrm{ch}(\varphi)$. We denote by $\Delta G$ the difference in Gibbs free-energy catalyzed in every reaction (every $2\pi$--cycle; Fig.~\ref{fig:fig1}a). $\Delta G$ is directly associated with the total chemical affinity in Michaelis-Menten cycles~\cite{Avanzini2019Dec}. In general, passive enzymes tend to lower energy barriers without altering reaction equilibria. In contrast, fueled-enzymes can use free-energy to perform a different work or drive a different process ``uphill''~\cite{Chatzittofi2025May}.

Figure~\ref{fig:fig1}(b) depicts an example of this reaction potential, which captures the binding of a substrate molecule by the enzyme and is associated with crossing of the energy barrier $E_\mathrm{b}$. Then, the catalytic reaction takes place and converts a substrate into a product molecule, which is associated with overcoming a second energy barrier $E_\mathrm{r}$. 
To include the enzymatic configurational degrees of freedom, we denote by $\p{R}_1$ and $\p{R}_2$ the position coordinates of the two enzymatic sub-units (Fig.~\ref{fig:fig1}c) and incorporate a contribution to the enzymatic potential energy $U^\mathrm{e}(\varphi, \p{R}_1, \p{R}_2) = V_\mathrm{ch}(\varphi) + V_m(\varphi, |\p{R}_1 - \p{R}_2|)$ which mechano-chemically couples the conformational dynamics to the catalytic process (its specific form will be discussed below).

In our setup, the enzyme is embedded in a bath of $N$ crowder particles whose three-dimensional positions are described by $\left\{\p{Y}_1, \cdots ,\p{Y}_N\right\}$ (Fig.~\ref{fig:fig1}c). The bath particles do not participate in the chemical reaction and are introduced to capture the mechanical features of the macromolecular environment. 
The chemical activity and all the mechanical interactions are governed by the total potential energy $\mathcal{U}(\varphi,\p{R}_1,\p{R}_2,\p{Y}_1,\cdots,\p{Y}_N)= U^\ee(\varphi,\p{R}_1,\p{R}_2) + \sum_{i=1}^2\sum_{a=1}^N U^{\ee\cc}_{i}(\p{R}_i,\p{Y}_a)$, where 
$U^{\ee\cc}_{i}(\p{R}_i,\p{Y}_a)=U^{\ee\cc}_{i}(|\p{Y}_a-\p{R}_i|)$ represent the steric coupling between the enzyme subunits and the crowder particles. 
The time-dependent probability distribution for these $3N+7$ degrees of freedom $\cP(\varphi,\p{R}_1,\p{R}_2,\p{Y}_1, \cdots ,\p{Y}_N)$ satisfies the Fokker-Planck equation
\begin{align}
    \partial_t \cP&= \partial_\varphi \left[M_\varphi  \left(k_\mathrm{B}T \partial_\varphi \cP + P\partial_\varphi \mathcal{U}\right)\right] \nonumber\\
    &+ \sum_{i,j=1}^2\bnabla_{\p{R}_i}\cdot \left[\p{M}^{\ee\ee}_{ij} \cdot \left(k_\mathrm{B} T \bnabla_{\p{R}_j} \cP + \cP {\bnabla}_{\p{R}_j} \mathcal{U}\right)\right]\nonumber\\
    &+\sum_{i=1}^2 \sum_{a=1}^N \bnabla_{\p{R}_i} \cdot \left[\p{M}^{\ee \cc}_{ia}\cdot \left(k_\mathrm{B} T \bnabla_{\p{Y}_a} \cP + \cP\bnabla_{\p{Y}_a}\mathcal{U}\right)\right]\nonumber\\
    & +\sum_{a=1}^N \sum_{i=1}^2 \!\bnabla_{\p{Y}_a} \cdot \left[\p{M}^{\cc\ee}_{ai} \cdot \left(k_\mathrm{B} T \bnabla_{\p{R}_i} \cP + \cP \bnabla_{\p{R}_i}\mathcal{U} \right) \right] \nonumber\\
    &+\sum_{a,b=1}^N\bnabla_{\p{Y}_a} \cdot \left[ \p{M}^{\cc\cc}_{ab} \cdot \left(k_\mathrm{B} T \bnabla_{\p{Y}_b}\cP + \cP \bnabla_{\p{Y}_b}\mathcal{U}\right)\right].
\end{align}
Here, $M_\varphi$ is the mobility associated with the reaction coordinate, $\p{M}^{\ee \ee}$ is the mobility tensor for the dynamics of the enzyme-subunits, $\p{M}^{\ee\cc}$ and $\p{M}^{\cc\ee}$ are the enzyme-crowder cross mobilities, and $\p{M}^{\cc \cc}$ represents the mobility tensor of the crowder particles when they interact with each other. For simplicity, we have neglected crowder-crowder interactions. 

To perform the coarse-graining that allows us to isolate the effective mechano-chemical dynamics of the enzyme, we integrate out the degrees of freedom associated with the crowder particles, as well as the center of mass of the enzyme $\p{R} \equiv (\p{R}_1+\p{R}_2)/2$, to obtain an equation for the marginal $\rho_\mathrm{e}(\varphi, \pell)=\int d^3\p{R} d^3\p{Y}_1 \cdots d^3\p{Y}_N \, \cP(\varphi,\p{R}_1=\p{R}-\frac12 \pell,\p{R}_2=\p{R}+\frac12 \pell,\p{Y}_1, \cdots,\p{Y}_N)$, where $\pell \equiv \p{R}_2-\p{R}_1$ describes the internal conformational degrees of freedom of the enzyme. This procedure leads to a hierarchy of coupled equations between different many-body distributions functions, and needs to be supplemented with an appropriate closure approximation scheme. After completing these steps, we obtain the following expression for the effective dynamics (see Supporting Information for details)
\begin{align}\label{eq:afterclo-3}
    \partial_t \rho_\ee &=\partial_\varphi \left[M_\varphi \big(k_\mathrm{B}T \partial_\varphi \rho_\ee + \rho_\ee\partial_\varphi U^\ee\big)\right]\nonumber\\&+\bnabla_{\pell}\cdot\left[\p{M} \cdot \Big(k_\mathrm{B} T \bnabla_{\pell}\rho_\ee + \rho_\ee \big(\bnabla_{\pell}U^\ee+\kappa \phi_{\cc}  \pell \big)\Big)\right],
\end{align}
to the lowest order, where $ \p{M} \equiv \p{M}^{\ee\ee}_{11}+\p{M}^{\ee\ee}_{22}-\p{M}^{\ee\ee}_{12}-\p{M}^{\ee\ee}_{21}$, and we have used a microscopic length scale $L_0$, which represents the geometric features of the enzyme, to define an effective crowder volume fraction $\phi_{\cc} \equiv \frac{4\pi}{3}\rho_{\cc} L_0^3$, with $\rho_\mathrm{c}$ being the uniform density of the crowders far from the enzyme. The effective fluctuation-induced {\em stiffness} $\kappa$ is defined as
\begin{align}
    \kappa&= \frac{k_\mathrm{B}T}{4\pi L_0^3} \int d^3{\p{Y}} \, \Big\{ e^{-\beta [U_1^{\ee\cc}(\p{Y})+U_2^{\ee\cc}(\p{Y})]}\nonumber\\ &\bnabla_{\p{Y}} \left[\beta U_1^{\ee\cc}(\p{Y})\right] \cdot \bnabla_{\p{Y}} \left[\beta U_2^{\ee\cc}(\p{Y})\right] \Big\}, 
    \label{eq:kappa}
\end{align}
where $\beta=1/(k_\mathrm{B}T)$. We note that the stiffness depends on the average of the product of the forces exerted by the crowder particles on the two enzyme sub-units. Therefore, the stiffness $\kappa$ reflects the degree of coherence between the two fluctuating quantities. The interaction potentials between enzyme sub-unit $i$ ($i \in \{1,2\}$), which we consider to be a sphere of radius $a_i$, and the crowder particles can be taken to have a Lennard-Jones form
$U_i^{\ee\cc}(Y=|\p{Y}|) =4\varepsilon_i \left[\left(\frac{a_i}{Y}\right)^{12} - \left(\frac{a_i}{Y}\right)^{6}\right]$, where $\varepsilon_i$ is the corresponding energy scale, representing generic short-range interactions. To get an intuitive picture about the way this stiffness is controlled by the physical parameters of the system, we can use the following approximate closed-form expression for equal size sub-units
\begin{align}
\kappa \simeq& 12\sqrt{\pi}\,2^{1/6}\,\frac{a }{L_0^3}\,
\frac{\sqrt{k_\mathrm{B}T} \epsilon_1\epsilon_2}
{(\epsilon_1+\epsilon_2)^{3/2}}\,
e^{\beta(\epsilon_1+\epsilon_2)},
    \label{eq:kappa-2}  
\end{align}
which is valid in a regime where the fluctuations around the minimum of the interaction are relatively small (see Supporting Information for details and the corresponding expression for the case of unequal sub-units). Finally, we can approximate the corresponding mobilities at the lowest order in the far-field limit $\p{M} = \left[(6\pi\eta a_1)^{-1} +(6\pi\eta a_2)^{-1}\right] \p{I} + \mathcal{O}(a/\ell)$. 

We now recast Eq.~\eqref{eq:afterclo-3} into an equation for the time evolution of the joint probability distribution for the enzyme extension and reaction coordinate $P_\mathrm{e}(\varphi,\ell)=4 \pi \ell^2 \rho_\ee(\varphi,\ell)$, which is normalized as $\int_{0}^{2 \pi} d \varphi   \int_{0}^{\infty} d \ell  \, P_\ee(\varphi,\ell) = 1$. The time evolution equation reads
\begin{align}\label{eq:twofp}
    &\partial_t P_\ee = \partial_\varphi \left\{ M_\varphi \big[k_\mathrm{B}T \partial_\varphi P_\ee + P_\ee\partial_\varphi U^\ee(\varphi,\ell)\big]\right\} \nonumber \\
    &+\partial_\ell \left\{ M_\ell \left[ k_\mathrm{B}T \partial_\ell P_\ee + P_\ee \left(\partial_\ell U^\ee(\varphi,\ell) + \kappa \phi_{\cc} \ell - \frac{2 k_\mathrm{B}T}{\ell} \right)\right]\right\},
\end{align}
where $M_\ell = \left[(6\pi\eta a_1)^{-1} +(6\pi\eta a_2)^{-1}\right]$. Equation~\eqref{eq:twofp} is the central result of this work. It elucidates the cross-talk between the conformational state of the enzyme and the reaction coordinate $\varphi$, while retaining the crowding dependence as an explicit mechanical force. We note that including crowder-crowder interaction can change the $\phi_{\cc}$ dependence of the interaction term, and the linear dependence in Eq.~\eqref{eq:twofp} is to be regarded as the first term in a Virial expansion.

As anticipated above, the enzymatic interaction potential $U^\mathrm{e}(\phi,\ell)=V_\mathrm{ch}(\varphi) + V_\mathrm{m}(\varphi,\ell)$ is composed of a purely chemical term $V_\mathrm{ch}(\varphi)= -\frac{\Delta G}{2\pi}\,\varphi + V_\mathrm{p}(\varphi)$ that drives the reaction coordinate down an energy landscape made by tilting a periodic potential $ V_\mathrm{p}(\varphi)$ by an excess free energy input $\Delta G$, and a mechano-chemical component $V_\mathrm{m}(\varphi,\ell)= \frac{k}{2}\,[\ell-L(\varphi)]^2$, taken here to have the form of a harmonic spring of stiffness $k$ with a $\varphi$-dependent rest length $L(\varphi)$. Putting all the components together, we have
\begin{align}
   U^\mathrm{e}(\phi,\ell) = -\frac{\Delta G}{2\pi}\,\phi + V_\mathrm{p}(\phi)+\frac{k}{2}\,[\ell-L(\phi)]^2. \label{eq:Ue-def}
\end{align}
Recall that the form of $V_\mathrm{p}(\varphi)$ determines the energy barriers $E_\mathrm{b}$ and $E_\mathrm{r}$ as shown in Fig.~\ref{fig:fig1}(b). Since the reaction coordinate is periodic, we choose $V_\mathrm{p}(\phi)= -v_1 \cos(\varphi+\delta_1) -v_2 \cos(2\varphi + \delta_2)$, representing the first two terms in a harmonic expansion, as an example. For the purpose of illustration, we also choose the phase shifts to be $\delta_1 = 0 $ and $\delta_2=\mathrm{arcsin}(\Delta G/4v_2\pi)$, such that the beginning and end of the reaction are located in multiples of $2\pi$. We note that at least two harmonics are required to get a landscape with two minima per cycle: we take the first minimum to correspond to the state of an unbounded (free) enzyme and the second minimum to when the enzyme binds a substrate molecule. 

In the mechanical coupling interaction term, $L(\varphi)$ serves as the rest-length of the enzyme. It depends on the reaction coordinate, allowing us to model the catalytic-induced conformational displacements. For consistency, $L(\varphi)$ must also be a $2\pi$-periodic function, guaranteeing that the average size of the enzyme returns to its original value after one cycle. Accordingly, we choose $L(\varphi)=L_0 [1 + \lambda \cos(\varphi+\delta)]$ where $L_0$ is the average rest-length of the enzymes and $\lambda$ is the maximum amplitude of the relative conformational change. With this simple choice, the slow dynamics of the enzyme undergoes a single expansion and contraction during a full catalytic cycle. The phase shift $\delta$ is a parameter that indicates the sequence of oscillations relative to the phase dynamics. For example, if $\delta = \pi/2$ the enzyme will first contract and then expand. 

For our analysis, we are interested in the impact of the crowder volume fraction $\phi_{\cc}$ on the dynamics of $\varphi$. Using separation of time-scales and assuming that the enzyme is stiff ($M_\varphi \Delta G \ll M_\ell k$), the spatial dynamics can be enslaved to the reaction coordinate (see Supporting Information for details). Note that this is a quasi-stationary elimination of the conformational coordinate, rather than an equilibrium approximation for the full catalytic cycle, which remains driven. We therefore project onto a one-dimensional description in which $\varphi$ is governed by the modified effective non-equilibrium free energy
\begin{align}
    V_\mathrm{eff}(\varphi) = V_\mathrm{ch}(\varphi)- 2k_\B T \ln L(\varphi) + \frac{1}{2}\kappa \phi_{\cc} L(\varphi)^2, \label{eq:Veff_phi}
\end{align}
which contains the crowder density. The second term is an entropic contribution associated with the rotational measure of the relative coordinate and is independent of crowding. The last term is the crowding-induced potential of mean force evaluated along the enzyme's preferred length $L(\varphi)$. In this sense, the construction is related to transfer-free-energy descriptions of crowding~\cite{ZhouRivasMinton2008,HuJiangRajagopalan2007,Minton2013StericChemical}, but it is embedded in a driven cyclic coordinate rather than in an equilibrium reaction scheme.

\begin{figure*}[t]
\centering\includegraphics[width=1.0\linewidth]{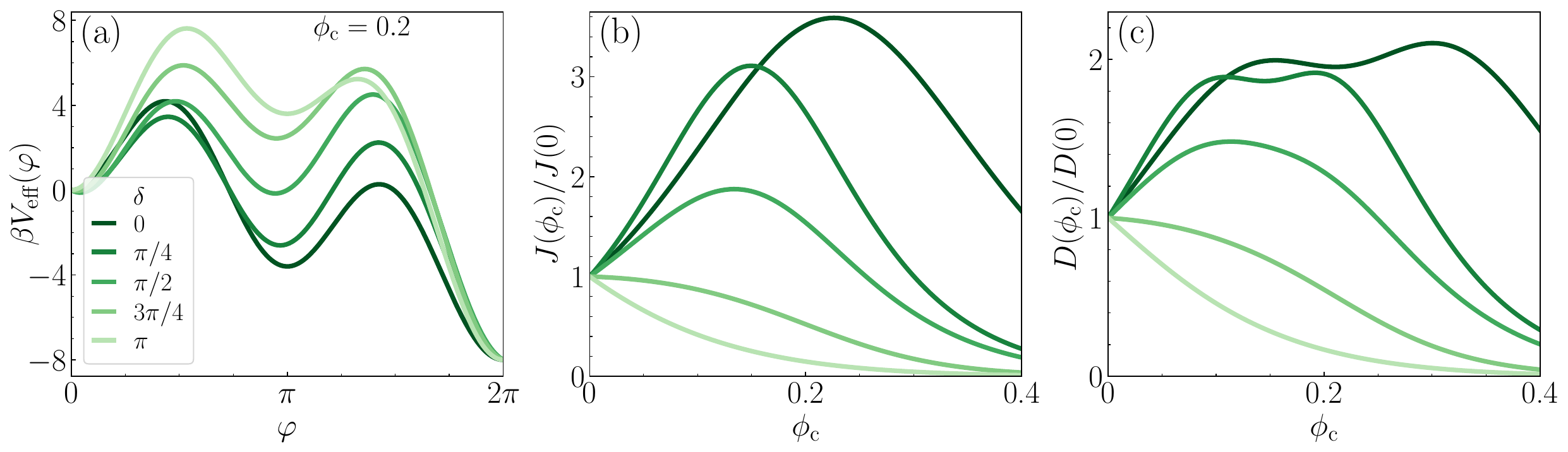}
\caption{Interplay between effective chemical non-equilibrium free energy, crowding and enzymatic activity. (a) Effective potentials that govern the stochastic dynamics for different $\delta$ and for fixed $\phi_{\cc}=0.2$. (b) The monotonic and non-monotonic response in the activity of the enzyme described by $J(\phi_{\cc})$ as function of dimensionless protein concentration and for different values of $\delta$. (c) The corresponding fluctuations of the current $J$. For these results the parameters are: $\beta \Delta G = 8$, $\beta v_1=2$, $\beta v_2=3$, $\lambda=0.1$ and $\beta \kappa L_0^2=100$. The choice of $v_1$ implies that the first and second minimum of $V_{\textrm{ch}}(\varphi)$ for $\phi_{\cc}=0$, are at the same energy level. See Supporting Information for further details about the dimensionless parameters.}
\label{fig:fig3}
\end{figure*}

As shown in Fig.~\ref{fig:fig3}(a), in this description the cycle affinity $\Delta G$ remains unchanged, and only the periodic part of the dynamics, namely the barrier structure around the cycle, is affected. It is worth emphasizing that this renormalization of the energy barriers is mathematically equivalent to modifying the enzyme's kinetic rates in a discrete-state description, through momentum conservation and the local detailed-balance condition~\cite{PhysRevResearch.6.L022044,Cocconi2025Jun}. For example, the usual detailed-balance condition relates forward ($k_+$) and backward ($k_-$) reaction rates to the local energy difference $\Delta \mu$, i.e. $k_+/k_- = e^{\beta \Delta \mu}$. In the presence of the effective fluctuation-induced force-dipole, the work performed during each conformational step must also be included~\cite{PhysRevResearch.6.L022044}. Hence, the local detailed balance is modified as $k_+(\phi_{\cc})/k_-(\phi_{\cc}) = e^{\beta \Delta \mu+c \beta \kappa \phi_{\cc} L_0^2/2}$, where $c$ is a numerical prefactor of order unity. See the Supporting Information for a reformulation of this result in the framework of the Michaelis-Menten kinetics as developed in the form of a Markov process.

A closely related theoretical study by Skora et al.~\cite{Skora2021ConformationCrowding} represents conformation-changing enzymes as fluctuating dumbbells, and finds that inert crowders promote the smaller closed state, thereby reducing activity. The present construction differs in two respects that are important for the results presented here. Firstly, the reaction is an explicitly driven cycle, which implies that the model remains thermodynamically consistent far from equilibrium. Secondly, the bare washboard potential $V_\mathrm{ch}(\varphi)$ supplies phase-dependent chemical barriers. Consequently, the same compressive load can lower a barrier when the enzyme is relatively expanded in the metastable state and more compact near the transition state, while raising another barrier elsewhere in the cycle. This barrier redistribution is the mechanism that allows both monotonic and non-monotonic responses within the same minimal model.

The effective one-dimensional picture allows us to analytically calculate the average reaction rate $J=\lim_{t\to \infty}\langle\varphi\rangle/2\pi t$ and the diffusion coefficient of the process $D = \lim_{t\to \infty}(\langle\varphi^2\rangle-\langle\varphi\rangle^2)/4\pi^2 t $, with the latter being a measure of the variability in the catalytic rate~\cite{PhysRevLett.87.010602}. Here, $\langle ... \rangle$ indicates the long-time average performed over the trajectory of the dynamics. The normalization by $2\pi$ accounts for the fact that full reactions are associated with $2\pi$ increments in $\varphi$. Figures~\ref{fig:fig3}(b) and \ref{fig:fig3}(c) show $J$ and $D$, respectively, as functions of the crowder concentration. For $\delta =0$ and $\pi/2$, the flux $J$ is maximized at intermediate values of $\phi_{\cc}$, while for $\delta = 3\pi/4$ and $\pi$ both the current and diffusivity decrease monotonically. Thus the sign of the crowding response is controlled by the phase relation between the conformational expansion and the chemical barriers, and not by changing the nature of the crowder. This provides a minimal route to understanding the diversity of enzyme-crowding trends observed experimentally~\cite{ZhouRivasMinton2008,RastogiChowdhury2021,Dindo2025Jan}.

From a thermodynamic point of view, the average metabolic activity in our model is directly related to the current in the only out-of-equilibrium reaction cycle available. Utilizing the framework of non-equilibrium stochastic thermodynamics~\cite{Seifert2019Mar,Cocconi2020Nov,RevModPhys.97.015002}, it is therefore straightforward to connect the entropy production rate $\dot \sigma$ of a single enzyme to its contribution to metabolic activity. In particular, at steady state, the enzymatic reactions dissipate $\Delta G$ at a rate equal to the flux $J$. The energy dissipated by the enzyme into the environment is thus given by the canonical form $T \dot \sigma = J(\phi_{\cc},\Delta G) \Delta G$~\cite{RevModPhys.48.571}. Since $J$ is a nonlinear function of the crowder density, the entropy production rate inherits the monotonic or non-monotonic crowding dependence, as shown in Fig.~\ref{fig:fig3}(b).

\begin{figure*}[t]
\centering\includegraphics[width=1.0\linewidth]{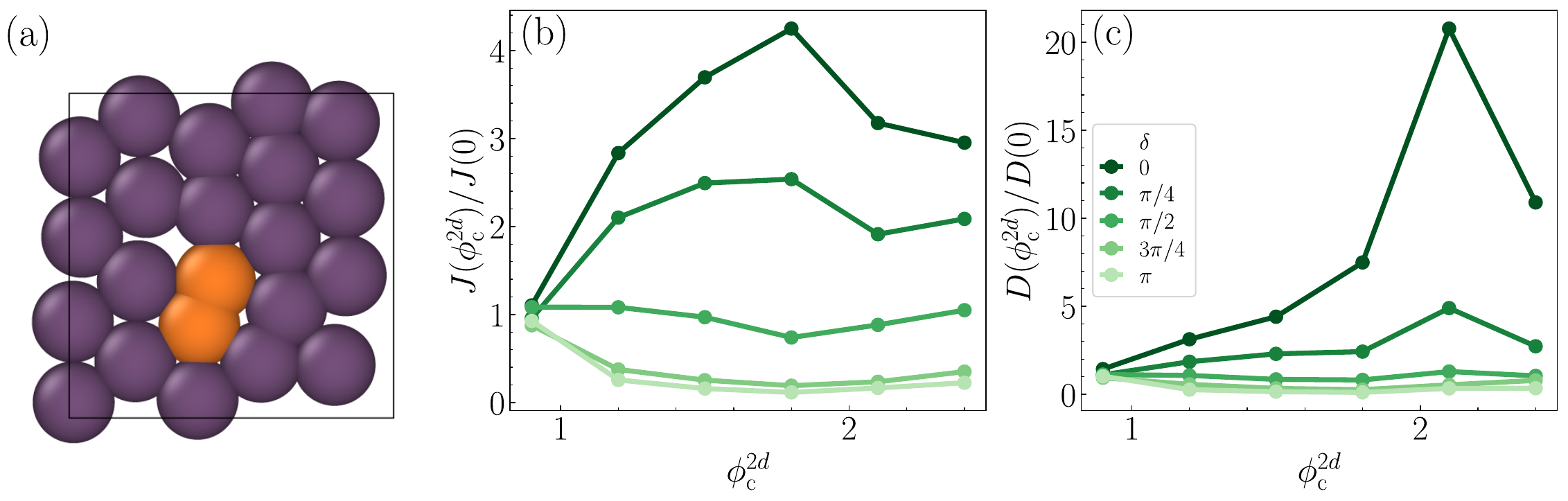}
\caption{Coarse-grained simulations of enzymes under different crowding conditions. 
We use periodic boundary conditions with area fraction greater than the circle packing fraction and a purely repulsive interaction potential to mimic crowding in the bulk of a condensate.
(a) Snapshot of dumbbell enzyme (orange) surrounded by crowder particles (purple).
(b) Measurements of the enzyme's catalytic rate, as a function of the background density, for different phase shifts $\delta$. 
(c) Measurements of the variability of the enzyme's catalytic rate, as a function of the background density, for different phase shifts $\delta$. Here $\phi_\mathrm{c}^{2d}$ corresponds to the area fraction.
See Supporting Information for details about the parameters.}
\label{fig:fig4}
\end{figure*}

Our analytical results are complemented by simulations, in which we numerically integrate the dynamics of a dumbbell enzyme surrounded by a bath of crowder particles in two dimensions. A snapshot of the setup is shown in Fig.~\ref{fig:fig4}(a). The dynamics of the internal variable $\varphi$ is determined by $U^\mathrm{e}(\varphi,\p{R}_1, \p{R}_2)$, with the addition of white noise to represent internal fluctuations of the enzyme within its reaction cycle. The crowder particle and enzyme subunit positions evolve based on a soft volume-exclusion interaction (including between crowders), and are also subject to noise, representing random motion due to the influence of solvent.  The enzyme subunits are furthermore subject to a force along their axis, which is derived from $V_\mathrm{m}$. 
This force drives the enzyme subunit separation towards the phase-dependent rest-length given by $L(\varphi)$. As shown in Fig.~\ref{fig:fig4}(b) and Fig.~\ref{fig:fig4}(c), above a critical crowding density (area fraction,  $\phi_\mathrm{c}^{2d}\approx1$ ), we observe the same range of behaviors predicted by the analytical results, including both monotonic and non-monotonic variations in the current and diffusion of the catalytic cycle as the crowder density is increased, depending on the phase shift $\delta$. Below this crowding density, the enzyme does not significantly feel the influence of the crowder particles and the current remains approximately constant, being driven mainly by the fluctuations of the enzyme itself. 
Overall, while direct quantitative comparison is hindered by simplifying theoretical assumptions and numerical stability constraints on the choice of interaction potentials, our simulations validates the predicted phase-shift dependence of the functional response of current and diffusion to variations in crowder density.

In this work, we have formulated a mechanistic non-equilibrium description of enzymatic catalysis in which chemical and mechanical coordinates remain coupled in a thermodynamically consistent way. Starting from the many-body stochastic dynamics of an enzyme and passive crowders, coarse-graining yields an effective fluctuation-induced force-dipole acting on the enzyme's conformational coordinate. The work done against this force modifies local detailed-balance and distorts the effective non-equilibrium free-energy landscape governing the reaction coordinate. Projecting to the one-dimensional cycle then gives us direct access to both the reaction current and its fluctuations, in an analytically tractable manner within our theoretical framework.

The resulting picture builds on established concepts that have been put forward in previous studies of the effect of crowding on the activity of enzymes. Excluded-volume and transfer-free-energy theories explain how crowding can favor compact conformations or shift association and isomerization equilibria~\cite{Minton1981Oct,ZhouRivasMinton2008,HuJiangRajagopalan2007,Minton2013StericChemical}. Simulations and experiments further show that hydrodynamic interactions, slowed diffusion, soft nonspecific interactions, quinary contacts, and crowder identity can all be important~\cite{AndoSkolnick2010,Schneider2015ADHCrowding,Wilcox2020CitrateSynthase,CandottiOrozco2016}. Our contribution is to isolate a complementary channel: even a uniformly compressive load can enhance turnover if the enzyme is relatively expanded in a metastable chemical state and more compact near the ensuing transition state. Since the enzyme's preferred conformation changes around the cycle, the same load can assist one transition and hinder another. Other features, such as specific activating interactions or the interplay of excluded volume, diffusion, and viscosity, can be incorporated in more elaborate models.

On a conceptual level, the emphasis on a driven mechano-chemical cycle can have advantages when we aim to build thermodynamically consistent frameworks for non-equilbrium systems in other contexts. Tools such as insertion free energies and chemical potentials are powerful equilibrium notions for passive crowded mixtures, but their direct extension to active or strongly driven media can become protocol- and coarse-graining-dependent~\cite{Cisneros2026InsertionActiveFluid}. Here, the affinity $\Delta G$, the condition of local detailed-balance, and the catalytic current remain explicit, and consequently, the model can be driven far from equilibrium without requiring the full crowded medium to be assigned an equilibrium free energy. This sheds light on the broader problem of characterizing dense, strongly interacting active systems~\cite{Cocconi2025Jun} and opens the door to uncertainty-relation bounds on observables of interest~\cite{Seifert2019Mar,PhysRevResearch.6.L022044,PhysRevResearch.6.L042012,PhysRevLett.132.037101,Dechant2018Dec}.

Although Eq.~\eqref{eq:twofp} was derived from a microscopic model of passive proteins, the same reduced dynamics can describe other sources of mechanical forcing~\cite{Robertson1990Oct,Golestanian2008a,Golestanian2010}. For instance, an enzyme with unequally charged subunits in an external electric field would experience an induced force dipole along its principal axis. In that case the force would be proportional to the field and the charge difference rather than to the crowder density, but the phase-dependent barrier-redistribution mechanism would be unchanged.

Finally, simple mechanistic models can complement high-dimensional molecular simulation and machine-learning approaches for understanding and designing proteins~\cite{Lovelock2022Jun,Frank2024Oct}. The present framework identifies geometric and kinetic parameters, such as the phase $\delta$ between conformational oscillation and chemical barriers, that could be optimized to make an enzyme robust, inhibited, or activated by mechanical loads. It also provides a low-dimensional dynamical system in which transition-path and optimization algorithms can be used to search for crowding-sensitive catalytic designs~\cite{PhysRevX.13.041044,PhysRevResearch.2.033208,heller2024evaluation}.

\begin{acknowledgements}
The authors acknowledge discussions with Mirco Dindo and Paola Laurino.
The authors acknowledge support from the Alexander von Humboldt Foundation and Ernst Oppenheimer Fund, from the Max Planck School Matter to Life and the MaxSynBio Consortium, which are jointly funded by the Federal Ministry of Education and Research (BMBF) of Germany and the Max Planck Society and funding from EPSRC through grant EP/Z534766/1.
\end{acknowledgements}

\bibliography{bibliography,Golestanian}

\clearpage
\onecolumngrid   

\setcounter{equation}{0}
\setcounter{figure}{0}
\setcounter{table}{0}
\setcounter{section}{0}
\setcounter{page}{1}
\renewcommand{\theequation}{S\arabic{equation}}
\renewcommand{\thefigure}{S\arabic{figure}}
\renewcommand{\thetable}{S\arabic{table}}
\renewcommand{\thesection}{S-\Roman{section}}
\renewcommand{\thepage}{S\arabic{page}}

\renewcommand{\theHequation}{S\arabic{equation}}
\renewcommand{\theHfigure}{S\arabic{figure}}
\renewcommand{\theHtable}{S\arabic{table}}
\renewcommand{\theHsection}{S-\Roman{section}}

\begin{center}
  {\large\bfseries Mechano-chemical control of enzymatic activity in crowded environments}\\[0.4em]
  {\itshape Supporting Information: Theoretical framework and derivations}\\[1.2em]

  Michalis Chatzittofi,\textsuperscript{1,2,*}
  Luca Cocconi,\textsuperscript{1,3}
  Jakob Metson,\textsuperscript{1}
  and Ramin Golestanian\textsuperscript{1,4,\dag}\\[0.8em]

  {\small\itshape
  \textsuperscript{1}Max Planck Institute for Dynamics and Self-Organization (MPI-DS), D-37077 G\"ottingen, Germany\\
  \textsuperscript{2}DAMTP, Centre for Mathematical Sciences, University of Cambridge, Cambridge CB3 0WA, United Kingdom\\
  \textsuperscript{3}TCM Group, Cavendish Laboratory, University of Cambridge, Cambridge CB3 0US, United Kingdom\\
  \textsuperscript{4}Rudolf Peierls Centre for Theoretical Physics, University of Oxford, Oxford OX1 3PU, UK\\[0.6em]
  }
  {\small
  \textsuperscript{*}\texttt{mc2623@cam.ac.uk}\quad
  \textsuperscript{\dag}\texttt{ramin.golestanian@ds.mpg.de}\\[0.4em]
  (Dated: \today)}
\end{center}
\vspace{1em}

\section{Coarse graining of a minimal enzyme model with mechano-chemical coupling}\label{sec:coarse-grain}

Here, we present our theoretical framework, and the details of the analysis that leads to our main findings, for the interplay between the catalytic efficiency and the mechanical coupling with the crowded environment of an enzyme. Our minimal model for the catalytic activity of an enzyme and its interaction with a crowded environment, requires a reaction coordinate $\varphi \in \mathbb{R}$ that represents the nonequilibrium chemical cycle, the enzyme (e) subunit positions $\p{R}_i \in \mathbb{R}^3$ (for $i \in \{1,2\}$), and the positions of $N$ crowder (c) molecules (e.g. proteins, etc) $\p{Y}_a \in \mathbb{R}^3$ (for $a \in \{1,2,...,N\}$). We assume that the system has translational and rotational symmetries, and use them to simplify the expressions when possible. For integration over the positions, we will use the shorthand notation $\int d^3 \p{Y} \,G(\p{Y})\equiv \int_{\p{Y}}G(\p{Y})$ etc.

The chemical activity and all the mechanical interactions are governed by the total potential energy $\mathcal{U}$, which we choose to have the following form
\begin{align}
\mathcal{U}  = U^\ee(\varphi,\p{R}_1,\p{R}_2) + \sum_{i=1}^2\sum_{a=1}^N U^{\ee\cc}_{i}(\p{R}_i,\p{Y}_a).
\end{align}
Here $U^\ee(\varphi,\p{R}_1,\p{R}_2)=U^\ee(\varphi,|\p{R}_2-\p{R}_1|)$ governs the dynamics of the enzyme via a mechanochemical coupling of the subunit positions to the reaction coordinate, and $U^{\ee\cc}_{i}(\p{R}_i,\p{Y}_a)=U^{\ee\cc}_{i}(|\p{Y}_a-\p{R}_i|)$ represent the mechanical coupling between the enzyme subunits and the crowder particles. Note that the enzyme sub-units can be different from one another while the crowder particles are taken to be indistinguishable. For simplicity, we have also neglected the (direct) interactions between the crowder particles, which amounts to treating the crowder suspension as being in the dilute regime (see below).

The probability distribution for these degrees of freedom $\cP(\varphi,\p{R}_1,\p{R}_2,\p{Y}_1,...,\p{Y}_N)$ satisfies the Fokker-Planck equation
\begin{align}
    \partial_t \cP&= \partial_\varphi \left[M_\varphi  \left(k_\mathrm{B}T \partial_\varphi \cP + P\partial_\varphi \mathcal{U}\right)\right] + \sum_{i,j=1}^2\bnabla_{\p{R}_i}\cdot \left[\p{M}^{\ee\ee}_{ij} \cdot \left(k_\mathrm{B} T \bnabla_{\p{R}_j} \cP + \cP {\bnabla}_{\p{R}_j} \mathcal{U}\right)\right]\nonumber\\
    &+\sum_{i=1}^2 \sum_{a=1}^N \bnabla_{\p{R}_i} \cdot \left[\p{M}^{\ee \cc}_{ia}\cdot \left(k_\mathrm{B} T \bnabla_{\p{Y}_a} \cP + \cP\bnabla_{\p{Y}_a}\mathcal{U}\right)\right]
     +\sum_{a=1}^N \sum_{i=1}^2\bnabla_{\p{Y}_a} \cdot \left[\p{M}^{\cc\ee}_{ai} \cdot \left(k_\mathrm{B} T \bnabla_{\p{R}_i} \cP + \cP \bnabla_{\p{R}_i}\mathcal{U} \right) \right] \nonumber\\
    &+\sum_{a,b=1}^N\bnabla_{\p{Y}_a} \cdot \left[ \p{M}^{\cc\cc}_{ab} \cdot \left(k_\mathrm{B} T \bnabla_{\p{Y}_b}\cP + \cP \bnabla_{\p{Y}_b}\mathcal{U}\right)\right].
\end{align}
Here, $M_\varphi$ is the mobility associated with the reaction coordinate, $\p{M}^{\ee \ee}$ is the mobility tensor for the dynamics of the enzyme-subunits, $\p{M}^{\ee\cc}$ and $\p{M}^{\cc\ee}$ are the enzyme-crowder cross mobilities, and $\p{M}^{\cc \cc}$ represents the mobility tensor of the crowder particles when they interact with each other. The mobilities are, in principle, position-dependent, and divergence-free, due to the incompressibility of the solvent. Moreover, they are subject to Onsager reciprocity relations \cite{DeGroot2013}. 

We are interested in deriving the effective dynamics for the reaction coordinate in the presence of a fluctuating mechanical environment. To this end, we seek to integrate out all the crowder degrees of freedom and focus on the reduced description that governs the enzyme marginal probability distribution 
\begin{align}
    \rho_\ee^{(2)}(\varphi,\p{R}_1,\p{R}_2) = \int_{\p{Y}_1, \cdots, \p{Y}_N} \cP(\varphi,\p{R}_1,\p{R}_2,\p{Y}_1,...,\p{Y}_N).
\end{align}
However, performing the integrals in the full Fokker-Planck equation generically produces an infinite hierarchy of equations (the BBGKY hierarchy), involving higher-order distributions. In this case, the joint three-body enzyme-crowder distribution function defined as
\begin{align}
\rho_{\ee\cc}^{(3)}(\varphi,\p{R}_1,\p{R}_2, \p{Y}) = \int_{\p{Y}_1, \cdots, \p{Y}_N} \delta(\p{Y}_1-\p{Y}) \,\cP(\varphi,\p{R}_1,\p{R}_2,\p{Y}_1,...,\p{Y}_N),
\end{align}
will feature in 
\begin{align}
\sum_{i,j=1}^2\int_{\p{Y}_1, \cdots, \p{Y}_N} \bnabla_{\p{R}_i} \cdot \left[\p{M}^{\ee\ee}_{ij} \cdot (\cP \bnabla_{\p{R}_j}\mathcal{U})\right]=\sum_{i,j=1}^2\bnabla_{\p{R}_i} \cdot \left[ \p{M}^{\ee\ee}_{ij} \cdot\left(\rho_\ee^{(2)} \bnabla_{\p{R}_j}U^\ee + N \int_{\p{Y}} \rho_{\ee\cc}^{(3)} \bnabla_{\p{R}_j} U_j^{\ee\cc}(\p{R}_j,\p{Y}) \right)\right], 
\end{align}
and
\begin{align}
    \sum_{i=1}^2 \bnabla_{\p{R}_i} \cdot \sum_{a=1}^N  \int_{\p{Y}_1, \cdots, \p{Y}_N} \p{M}_{ia}^{\ee\cc} \cdot \cP \bnabla_{\p{Y}_a}\mathcal{U}=N\sum_{i=1}^2 \bnabla_{\p{R}_i} \cdot \int_{\p{Y}} \p{M}_{i}^{\ee\cc} \cdot \rho_{\ee\cc}^{(3)} \, \bnabla_{\p{Y}} \left(\sum_{k=1}^2  U^{\ee\cc}_k(\p{R}_k,\p{Y})\right),
\end{align}
where $\p{M}^{\ee\cc}_i = \p{M}^{\ee\cc}_i(\p{R}_i,\p{Y})$. Collecting all the relevant terms, we find the following dynamical equation
\begin{align}\label{eq:beforeclo}
    \partial_t \rho_\ee^{(2)} &=\partial_\varphi \left[M_\varphi \big(k_\mathrm{B}T \partial_\varphi \rho_\ee^{(2)} + \rho_\ee^{(2)}\partial_\varphi U^\ee\big)\right]+ \sum_{i,j=1}^2 \bnabla_{\p{R}_i}\cdot \left[ \p{M}^{\ee\ee}_{ij} \cdot \big(k_\mathrm{B} T \bnabla_{\p{R}_j}\rho_\ee^{(2)} + \rho_\ee^{(2)}\bnabla_{\p{R}_j}U^\ee \big)\right]\nonumber\\
    &+ N\sum_{i,j=1}^2 \int_{\p{Y}} \bnabla_{\p{R}_i}\cdot \left[ \p{M}^{\ee\ee}_{ij} \cdot  \rho_{\ee\cc}^{(3)}\bnabla_{\p{R}_j}  U^{\ee\cc}_j(\p{R}_j,\p{Y}) \right]\nonumber\\
    &+N\sum_{i=1}^2 \int_{\p{Y}} \bnabla_{\p{R}_i}\cdot \left[\p{M}^{\ee\cc}_{i} \cdot \Big(k_\mathrm{B}T \bnabla_\p{Y}\rho_{\ee\cc}^{(3)} + \rho_{\ee\cc}^{(3)}\bnabla_\p{Y}\big[U^{\ee\cc}_1(\p{R}_1,\p{Y})+U^{\ee\cc}_2(\p{R}_2,\p{Y})\big] \Big) \right].
\end{align}

To close the hierarchy of the equations, we use the following closure scheme
\begin{align}
\rho_{\ee\cc}^{(3)}(\varphi,\p{R}_1,\p{R}_2, \p{Y}) &\simeq \frac{\rho_{\cc}}{N}\,\rho_{\ee}^{(2)}(\varphi,\p{R}_1,\p{R}_2) \,\exp\left(-\beta \big[U^{\ee \cc}_1(\p{R}_1,\p{Y})+U^{\ee \cc}_2(\p{R}_2,\p{Y})\big]\right), 
\label{eq:closure_alignment}
\end{align}
where $\beta=1/(k_\mathrm{B}T)$, and $\rho_{\cc}$ is the (uniform) density of the crowders far from the enzyme. Equation \eqref{eq:closure_alignment} amounts to a separation of time scales corresponding to the case where the distribution of the crowders follows that of the enzyme in a quasi-stationary manner, which causes the third line in Eq.~\eqref{eq:beforeclo} to vanish. Inserting Eq. \eqref{eq:closure_alignment} in Eq. \eqref{eq:beforeclo} yields
\begin{align}\label{eq:afterclo}
    \partial_t \rho_\ee^{(2)} &=\partial_\varphi \left[M_\varphi \big(k_\mathrm{B}T \partial_\varphi \rho_\ee^{(2)} + \rho_\ee^{(2)}\partial_\varphi U^\ee\big)\right]\nonumber \\
    &+\sum_{i,j=1}^2 \bnabla_{\p{R}_i}\cdot \left[ \p{M}^{\ee\ee}_{ij} \cdot \left( k_\mathrm{B} T \bnabla_{\p{R}_j}\rho_\ee^{(2)} + \rho_\ee^{(2)} \Big(\bnabla_{\p{R}_j}U^\ee -\rho_{\cc} \int_{\p{Y}} e^{-\beta(U_1^{\ee\cc}+U_2^{\ee\cc})} \bnabla_{\p{Y}}U^{\ee\cc}_j\Big)\right)\right].
\end{align}

We now introduce the coordinate transformation $\p{R}=(\p{R}_1+\p{R}_2)/2$ and ${\pell} = \p{R}_2 - \p{R}_1$, or, conversely, $\p{R}_1=\p{R}-\frac12 \pell$ and $\p{R}_2=\p{R}+\frac12 \pell$, which builds on the symmetries of the problem. Similarly, for the gradients we have $\bnabla_{\p{R}_1} = \frac{1}{2}\bnabla_\p{R} - \bnabla_{\pell}$ and $\bnabla_{\p{R}_2} = \frac{1}{2}\bnabla_\p{R} + \bnabla_{\pell}$. Invoking the translational symmetry and rewriting everything in terms of the new coordinate, we find $U^\ee(\varphi,\p{R}_1,\p{R}_2)=U^\ee(\varphi,\pell)$, $U^{\ee\cc}_{1}(\p{R}_1,\p{Y})=U^{\ee\cc}_{1}(\p{Y}-\p{R}+\frac12 \pell)$, $U^{\ee\cc}_{2}(\p{R}_2,\p{Y})=U^{\ee\cc}_{2}(\p{Y}-\p{R}-\frac12 \pell)$. We now note that inside the $\p{Y}$ integrals, we could eliminate the dependence on $\p{R}$ by using the change of variable $\p{Y} \to \p{Y}-\p{R}$. Since the mobilities do not depend on $\p{R}$, we can now integrate out $\p{R}$ in Eq. \eqref{eq:afterclo}, and derive the dynamical equation for the following one-body distribution
\begin{align}
    \rho_\ee(\varphi,\pell) = \int_{\p{R}} \rho_\ee^{(2)}\left(\varphi,\p{R}_1,\p{R}_2\right) =\int_{\p{R}} \rho_\ee^{(2)}\left(\varphi,\p{R}-\frac12 \pell,\p{R}+\frac12 \pell\right),
\end{align}
which governs the internal degrees of freedom of the enzyme. Using the definitions 
\begin{align}
  U^{\ee\cc}(\p{Y},\pell) &\equiv U^{\ee\cc}_{1}(\p{Y}+\tfrac12 \pell)+U^{\ee\cc}_{2}(\p{Y}-\tfrac12 \pell),  \label{eq:U-def} \\
   u^{\ee\cc}(\p{Y},\pell) &\equiv U^{\ee\cc}_{1}(\p{Y}+\tfrac12 \pell)-U^{\ee\cc}_{2}(\p{Y}-\tfrac12 \pell), \label{eq:u-def} 
\end{align}
for effective potentials, and 
\begin{align}
    \p{M} \equiv \p{M}^{\ee\ee}_{11}+\p{M}^{\ee\ee}_{22}-\p{M}^{\ee\ee}_{12}-\p{M}^{\ee\ee}_{21},\label{eq:M-def}
\end{align}
for a combination of the mobilities, the resulting effective dynamics reads
\begin{align}\label{eq:afterclo-2}
    \!\!\partial_t \rho_\ee =\partial_\varphi \!\left[M_\varphi \big(k_\mathrm{B}T \partial_\varphi \rho_\ee + \rho_\ee\partial_\varphi U^\ee\big)\right]+\bnabla_{\pell}\cdot\left[ \p{M} \cdot \left(k_\mathrm{B} T \bnabla_{\pell}\rho_\ee + \rho_\ee \Big(\bnabla_{\pell}U^\ee+\frac{\rho_{\cc}}{2} \int_{\p{Y}} \!e^{-\beta U^{\ee\cc}(\p{Y},\pell)} \bnabla_{\p{Y}}u^{\ee\cc}(\p{Y},\pell)\Big)\right)\right],
\end{align}
where a boundary term has been eliminated. 
We now perform the following Taylor expansions 
\begin{align}
  U^{\ee\cc}(\p{Y},\pell) &= U^{\ee\cc}(\p{Y},\p{0}) +\tfrac12 \pell \cdot \bnabla_{\p{Y}} u^{\ee\cc}(\p{Y},\p{0})+O(\pell^2)=U^{\ee\cc}(\p{Y}) +\tfrac12 \pell \cdot \bnabla_{\p{Y}} u^{\ee\cc}(\p{Y})+\mathcal{O}(\pell^2)   \label{eq:U-exp-1} \\
   u^{\ee\cc}(\p{Y},\pell) &= u^{\ee\cc}(\p{Y},\p{0}) +\tfrac12 \pell \cdot \bnabla_{\p{Y}} U^{\ee\cc}(\p{Y},\p{0})+O(\pell^2)=u^{\ee\cc}(\p{Y}) +\tfrac12 \pell \cdot \bnabla_{\p{Y}} U^{\ee\cc}(\p{Y})+\mathcal{O}(\pell^2), \label{eq:u-exp-1} 
\end{align}
where the definitions $U^{\ee\cc}(\p{Y})\equiv U^{\ee\cc}(\p{Y},\p{0})$ and $u^{\ee\cc}(\p{Y}) \equiv u^{\ee\cc}(\p{Y},\p{0})$ have been incorporated. Inserting the Taylor expansion in the interaction term in Eq. \eqref{eq:afterclo-2}, we obtain
\begin{align}
    \int_{\p{Y}} \!e^{-\beta U^{\ee\cc}(\p{Y},\pell)} \bnabla_{\p{Y}}u^{\ee\cc}(\p{Y},\pell) &=\int_{\p{Y}} \!e^{-\beta U^{\ee\cc}(\p{Y})} \bnabla_{\p{Y}}u^{\ee\cc}(\p{Y})-\frac{\beta}{2} \int_{\p{Y}} \!e^{-\beta U^{\ee\cc}(\p{Y})} \big[\pell \cdot \bnabla_{\p{Y}} u^{\ee\cc}(\p{Y})\big] \bnabla_{\p{Y}} u^{\ee\cc}(\p{Y}) \nonumber \\
    &+\frac{\beta}{2} \int_{\p{Y}}  \!e^{-\beta U^{\ee\cc}(\p{Y})} \big[\pell \cdot \bnabla_{\p{Y}} U^{\ee\cc}(\p{Y})\big] \bnabla_{\p{Y}} U^{\ee\cc}(\p{Y})+\mathcal{O}(\pell^2).\label{eq:Y-int-2}
\end{align}
Invoking the rotational symmetry of the potentials, namely, $U^{\ee\cc}(\p{Y}) = U^{\ee\cc}(|\p{Y}|)$
and $u^{\ee\cc}(\p{Y}) =u^{\ee\cc}(|\p{Y}|)$, we find that Eq. \eqref{eq:Y-int-2} simplifies as follows
\begin{align}
    \int_{\p{Y}} \!e^{-\beta U^{\ee\cc}(\p{Y},\pell)} \bnabla_{\p{Y}}u^{\ee\cc}(\p{Y},\pell) &=
    \frac{1}{6} \beta \pell \int_{\p{Y}} \!e^{-\beta U^{\ee\cc}(\p{Y})} \big[\big(\bnabla_{\p{Y}} U^{\ee\cc}(\p{Y})\big)^2-\big(\bnabla_{\p{Y}} u^{\ee\cc}(\p{Y})\big)^2\big] + \mathcal{O}(\pell^2),\label{eq:Y-int-3}
\end{align}
which can also be rewritten as
\begin{align}
    \int_{\p{Y}} \!e^{-\beta U^{\ee\cc}(\p{Y},\pell)} \bnabla_{\p{Y}}u^{\ee\cc}(\p{Y},\pell) &=
    \frac{2}{3} \beta \pell \int_{\p{Y}} \!e^{-\beta [U_1^{\ee\cc}(\p{Y})+U_2^{\ee\cc}(\p{Y})]} \,\bnabla_{\p{Y}} U_1^{\ee\cc}(\p{Y}) \cdot \bnabla_{\p{Y}} U_2^{\ee\cc}(\p{Y}) + \mathcal{O}(\pell^2).\label{eq:Y-int-4}
\end{align}
Inserting the above results in Eq. \eqref{eq:afterclo-2}, we obtain the following expression for the effective dynamics
\begin{align}\label{seq:afterclo-3}
    \!\!\partial_t \rho_\ee =\partial_\varphi \!\left[M_\varphi \big(k_\mathrm{B}T \partial_\varphi \rho_\ee + \rho_\ee\partial_\varphi U^\ee\big)\right]+\bnabla_{\pell}\cdot\left[\p{M} \cdot \Big(k_\mathrm{B} T \bnabla_{\pell}\rho_\ee + \rho_\ee \big(\bnabla_{\pell}U^\ee+\kappa \phi_{\cc}  \pell \big)\Big)\right],
\end{align}
to the lowest order, where we have used a microscopic length scale $L_0$, which represents the geometric features of the enzyme (see below), to define an effective crowder volume fraction 
\begin{align}
    \phi_{\cc} \equiv \frac{4\pi}{3}\rho_{\cc} L_0^3. 
\end{align}
The effective fluctuation-induced {\em stiffness} $\kappa$ is defined as
\begin{align}
    \kappa= \frac{k_\mathrm{B}T}{ 4\pi L_0^3} \int_{\p{Y}} e^{-\beta [U_1^{\ee\cc}(\p{Y})+U_2^{\ee\cc}(\p{Y})]} \,\bnabla_{\p{Y}} \left[\beta U_1^{\ee\cc}(\p{Y})\right] \cdot \bnabla_{\p{Y}} \left[\beta U_2^{\ee\cc}(\p{Y})\right], 
    \label{eq:kappa}
\end{align}
and depends on the average of the product of the forces on the two enzyme sub-units. Therefore, the stiffness $\kappa$ reflects the degree of coherence between the two fluctuating quantities. The interaction potentials between enzyme sub-unit $i$ ($i \in \{1,2\}$), which we consider to be a sphere of radius $a_i$, and the crowder particles can be taken to have a Lennard-Jones form
\begin{align}
    U_i^{\ee\cc}(Y=|\p{Y}|) =4\varepsilon_i \left[\left(\frac{a_i}{Y}\right)^{12} - \left(\frac{a_i}{Y}\right)^{6}\right],
     \label{eq:U-LJ}
\end{align}
and thus,
\begin{align}
    U^{\ee\cc} &= \epsilon \left[ \left(\frac{Y_0}{Y}\right)^{12}-2\left( \frac{Y_0}{Y}\right)^{6}\right],\\
    u^{\ee\cc} &= \epsilon \left[ b_{12}\left(\frac{Y_0}{Y}\right)^{12}-2b_6\left( \frac{Y_0}{Y}\right)^{6}\right].\\
\end{align}
For compactness of the presentation, we define the following quantities
\begin{align}
    Y_0 &= 2^{1/6}\left( \frac{\epsilon_1 a_1^{12}+\epsilon_2 a_2^{12}}{\epsilon_1 a_1^{6}+\epsilon_2 a_2^{6}}\right)^{1/6},\\
    \epsilon &= \frac{(\epsilon_1 a_1^{6}+\epsilon_2 a_2^{6})^2}{\epsilon_1 a_1^{12}+\epsilon_2 a_2^{12}},\\
    b_{12} &= \frac{\epsilon_1 a_1^{12}-\epsilon_2 a_2^{12}}{\epsilon_1 a_1^{12}+\epsilon_2 a_2^{12}},\\
    b_{6} &= \frac{\epsilon_1 a_1^{6}-\epsilon_2 a_2^{6}}{\epsilon_1 a_1^{6}+\epsilon_2 a_2^{6}}.
\end{align}
We then define $y=Y/Y_0$ and we expand $y=1+\delta$. Hence, $U(\delta) \simeq U^{\ee\cc}(\delta)+\mathcal{O}(\delta^3)$ and $u(\delta) \simeq u^{\ee\cc}(\delta)+\mathcal{O}(\delta^4)$, such that
\begin{align}
    [U^{\prime}(\delta)]^2 &= 72^2 \epsilon^2 \delta^2,\\
    [u^{\prime}(\delta)]^2 &= 12^2\epsilon^2(b_{12}-b_6)^2-48\epsilon^2(b_{12}-b_6)(78b_{12}-42b_6)\delta + \left[ 4\epsilon^2(78b_{12}-42b_6)^2+48\epsilon^2(b_{12}-b_6)(546b_{12}-168b_6)\right]\delta^2.
\end{align}
Therefore, using the integral expression of $\kappa$, which can be written as
\begin{align}
    \kappa \simeq \frac{k_BT\, Y_0}{4 L_0^3} \int_{-\infty}^{\infty} d\delta \, (1+2 \delta +\delta^2)\,
    e^{-\beta U} \left[ U'^2 - u'^2 \right],
\end{align}
we obtain
\begin{align}
    \kappa \simeq\frac{Y_0}{12 L_0^3}\sqrt{\pi k_BT \epsilon} \, e^{\beta \epsilon}\left[ 36-300b_{12}^2+342b_{12}b_6-78b_6^2-72\beta\epsilon(b_{12}-b_6)^2\right].
\end{align}
For $a_1=a_2$ we recover Eq.~(5) in the main text.
Finally, we can approximate the corresponding mobilities at the lowest order in the far-field limit, which yields
\begin{align}
    \p{M} = \left[(6\pi\eta a_1)^{-1} +(6\pi\eta a_2)^{-1}\right] \p{I} + \mathcal{O}(a/\ell).\label{eq:M-2}
\end{align}
We now recast Eq. \eqref{seq:afterclo-3} into an equation for the time evolution of the joint probability distribution for the enzyme extension and reaction coordinate $P_\mathrm{e}(\varphi,\ell)$, which is defined by integrating over the orientation of $\pell$ as follows
\begin{align}
    P_\ee(\varphi,\ell) = \int_{\pell} \rho_\ee(\varphi,{\pell}) \delta(|\pell|-\ell)=4 \pi \ell^2 \rho_\ee(\varphi,\ell),
\end{align}
and is normalized as follows
\begin{align}
    \int_{0}^{2 \pi} d \varphi   \int_{0}^{\infty} d \ell  \, P_\ee(\varphi,\ell) = 1.
\end{align}
The time evolution equation reads
\begin{align}\label{seq:twofp}
    \partial_t P_\ee = \partial_\varphi \left\{ M_\varphi \big[k_\mathrm{B}T \partial_\varphi P_\ee + P_\ee\partial_\varphi U^\ee(\varphi,\ell)\big]\right\}+\partial_\ell \left\{ M_\ell \left[ k_\mathrm{B}T \partial_\ell P_\ee + P_\ee \left(\partial_\ell U^\ee(\varphi,\ell) + \kappa \phi_{\cc} \ell - \frac{2 k_\mathrm{B}T}{\ell} \right)\right]\right\},
\end{align}
where $M_\ell = \left[(6\pi\eta a_1)^{-1} +(6\pi\eta a_2)^{-1}\right]$. This is Eq.~(4) in the main text. The interaction potential $U^\mathrm{e}(\phi,\ell)$ is composed of a component $V_\mathrm{ch}(\varphi)$ that drives $\varphi$ down an energy landscape made by tilting a periodic potential $ V_\mathrm{p}(\varphi)$ by an excess free energy input $\Delta G$, and a mechanical component $V_\mathrm{m}(\varphi,\ell)$ that couples the two degrees of freedom, taken to have the form of a harmonic spring of stiffness $k$ with a $\varphi$-dependent rest length $L(\varphi)$, namely
\begin{align}
U^\mathrm{e}(\varphi,\ell) &= V_\mathrm{ch}(\varphi) + V_\mathrm{m}(\varphi,\ell),\\
V_\mathrm{ch}(\varphi) &= -\frac{\Delta G}{2\pi}\,\varphi + V_\mathrm{p}(\varphi),\\
V_\mathrm{m}(\varphi,\ell) &= \frac{k}{2}\,[\ell-L(\varphi)]^2.
\end{align}
Putting all the components together, we have
\begin{align}
   U^\mathrm{e}(\phi,\ell) = -\frac{\Delta G}{2\pi}\,\phi + V_\mathrm{p}(\phi)+\frac{k}{2}\,[\ell-L(\phi)]^2. \label{eq:Ue-def}
\end{align}

\section{Derivation of the projected reaction coordinate dynamics}

In this section, we study the impact of the fluctuation-induced interactions with the crowding environment on the chemical activity of an enzyme by solving Eqs.~\eqref{seq:twofp} and \eqref{eq:Ue-def}. Using the derived Fokker-Planck equation, we can write the corresponding Langevin dynamics as follows
\begin{align}
    \dot \ell &= -M_\ell k[\ell-L(\varphi)] - M_\ell \kappa \phi_{\cc} \ell +  \frac{2 M_\ell k_B T}{\ell}+ \sqrt{2M_\ell k_\mathrm{B}T } \,\xi_\ell,\\
    \dot \varphi &= -M_\varphi[V'_\mathrm{ch}(\varphi)-k[\ell-L(\varphi)]L'(\varphi)]+ \sqrt{2M_\varphi k_\mathrm{B}T } \,\xi_\varphi\label{eq:phi_lownoise},
\end{align}
where $\xi_\varphi$ and $\xi_\ell$ are white noise terms with zero mean and unit variance. We define $\delta \ell \equiv \ell-L(\varphi)$ such that $\delta \dot\ell=\dot \ell - L'(\varphi)\dot\varphi$, derive a perturbative expansion for the (deterministic) evolution of $\delta \dot\ell$ as follows
\begin{align}
    \delta \dot\ell = -M_\ell k\delta \ell \left( 1 + \frac{\kappa \phi_{\cc}}{k}\right) -L'(\varphi)\dot\varphi-M_\ell \kappa \phi_{\cc} L(\varphi) + \frac{2 M_\ell k_\mathrm{B} T}{L(\varphi)}-\frac{2 M_\ell k_\mathrm{B} T \delta \ell}{L(\varphi)^2}+\mathcal{O}(\delta \ell^2 / L^2).
\end{align}
Assuming a separation of time scales between the dynamics of $\varphi$ and $\delta\ell$, 
we can simplify the dynamics by solving $\delta\dot\ell=0$ at fixed $(\varphi,\dot\varphi)$ and substituting it into Eq.~\eqref{eq:phi_lownoise}. This yields
\begin{align}
    \dot \varphi = -M_\varphi\left[\frac{M_\varphi}{M_\ell} L'(\varphi)^2+\left( 1+ \frac{\kappa \phi_{\cc}}{k} + \frac{2k_\mathrm{B}T}{kL^2}\right) \right]^{-1}\left[V'_\mathrm{ch}(\varphi) \left(1+\frac{\kappa \phi_{\cc}}{k} + \frac{2k_\mathrm{B}T}{kL^2}\right)+\kappa \phi_{\cc} L(\varphi) L'(\varphi)-\frac{2 k_\mathrm{B} T L'(\varphi)}{L(\varphi)}\right].
\end{align}
Therefore, the deterministic part of the dynamics takes on the form
\begin{align}
    \dot \varphi = M(\varphi)\left[ -\partial_\varphi V_\mathrm{eff}(\varphi)\right].
\end{align}
Since we have assumed strong mechanochemical coupling (i.e. large $k$), the potential and the mobility can be further simplified, to yield
\begin{align}\label{eq:veff}
    V_\mathrm{eff}(\varphi) &\equiv V_\mathrm{ch}(\varphi) -2 k_\mathrm{B}T \ln(L(\varphi))+ \frac{1}{2} \kappa \phi_{\cc} L(\varphi)^2, \\
    M(\varphi) &\equiv \frac{M_\varphi}{1 + \frac{M_\varphi}{M_\ell} L'(\varphi)^2}.
\end{align}

For our particular choices of $V_\mathrm{ch}(\varphi)$ and $L(\varphi)$, namely,
\begin{align}
    V_\mathrm{ch}(\varphi)&= -\frac{\Delta G}{2\pi}\, \varphi - v_1 \cos(\varphi+\delta_1) -v_2 \cos(2\varphi + \delta_2), \\
    L(\varphi)&=L_0 \left[1 + \lambda \cos(\varphi+\delta)\right],
\end{align}
with the phase shifts given as $\delta_1 = 0 $ and $\delta_2=\mathrm{sin}^{-1}(\Delta G/4v_2\pi)$, we find that the effective potential becomes 
\begin{align}
    V_\mathrm{eff}(\varphi)&=-\frac{\Delta G}{2\pi}\,\varphi -\left[v_1 -\kappa \phi_{\cc} L_0^2 \lambda \cos\delta+ 2k_\mathrm{B} T  \lambda \cos\delta\right]\cos\varphi -\left[\kappa \phi_{\cc} L_0^2 \lambda \sin\delta- 2k_\mathrm{B} T \lambda \sin\delta\right] \sin\varphi \nonumber \\
    &-\left[v_2 - \frac{1}{4} \kappa \phi_{\cc} L_0^2 \lambda^2\cos(2\delta-\delta_2) \right]\cos(2\varphi+\delta_2) - \frac{1}{16\pi} \kappa \phi_{\cc} L_0^2 \lambda^2\sin(2\delta-\delta_2)\sin(2\varphi + \delta_2),
\end{align}
where $\kappa$ features prominently.

\section{Estimating the relevant parameters}
The Gibbs free energy bias $\Delta G$ that drives the system and the energy barriers $v_1$ and $v_2$ are typically of the order of $\Delta G\sim 0.1-10k_\mathrm{B} T$. For the energy barriers this is also the case. The typical size of an enzyme is roughly $L_0\sim 1-10 \mathrm{nm}$ and the fractional conformational change scale can be taken as $\lambda \sim 0.1$. Finally, the energy scale of the interaction potential is roughly $\varepsilon \sim 1-10k_\mathrm{B}T$. We consider identical subunits, and set $a \sim L_0$. Taking specific values of $L_0 = 2$nm and $\rho_{\cc}= 10 \mathrm{mM} = 10^{25}/\mathrm{m}^3$, we obtain $\phi_{\cc}\approx 0.4$, while using $\varepsilon/k_\mathrm{B}T = 1.2$, we obtain $\kappa L_0^2/k_\mathrm{B}T=100$ by performing the integral in Eq. \eqref{eq:kappa} using the definition of Eq. \eqref{eq:U-LJ}. For the plots in the main text, we vary $\phi_{\cc}$ in the range $0-0.4$.

\section{Varying the mechano-chemical coupling}

Another way to get an analytical solution is in the limit of the fast dynamics of the length i.e $M_\ell \gg 1$. The mechano-chemical potential of the enzyme is given by
\begin{align}
   U^\mathrm{e}_\mathrm{eff}(\varphi,\ell) = -\frac{\Delta G}{2\pi}\,\varphi + V_\mathrm{p}(\varphi)+\frac{k}{2}\,[\ell-L(\varphi)]^2+\frac{\kappa}{2}\phi_{\cc}\ell^2. \label{eq:Ue-eff-def}
\end{align}
Using the separation of time scales used above, 
we obtain the conditional probability for a given $\varphi$ to be 
\begin{equation}
    P_\mathrm{fast}(\ell|\varphi) = \mathcal{N}_\varphi \exp\left[ -\frac{1}{k_\mathrm{B}T}\left(\frac{k}{2}\,[\ell-L(\varphi)]^2+\frac{\kappa}{2}\phi_{\cc}\ell^2 - 2 k_\mathrm{B}T \ln\ell \right) \right]~,
\end{equation}
where we consider $\ell \in [0,\infty)$ to account for hard-core repulsion. To obtain the normalization constant, we use
\begin{align}
    \int_0^\infty d\ell \, P_\mathrm{fast}(\ell|\varphi) = 1,
\end{align}
which yields
\begin{align}
    \mathcal{N}_\varphi^{-1}=\frac{\sqrt{k_\mathrm{\mathrm{B}}T} e^{-\frac{1}{2}\beta k L^2}}{\sqrt{2}
   (\kappa \phi_{\cc}+k)^{5/2}}\left(\sqrt{\pi }
   e^{\frac{\beta k^2 L^2}{2 (\kappa \phi_{\cc}+k)}} \left[k^2 L^2 + k_\mathrm{B}T(\kappa \phi_{\cc}+k) \right]
   \left(\mathrm{erf}\left(\sqrt{\frac{\beta k^2 L^2}{2 (\kappa \phi_{\cc}+k)}}\right)+1\right)+ k L \sqrt{2k_\mathrm{\mathrm{B}}T (\kappa \phi_{\cc}+k)}\right). 
\end{align}
The full Fokker-Planck equation of the driven process can be integrated with respect to $\ell$ to obtain
    \begin{equation}
    \partial_t P_\varphi = \partial_\varphi M_\varphi\left[ k_\mathrm{B} T \partial_\varphi P_\varphi + \left( -\frac{\Delta G}{2\pi} + \partial_\varphi V_\mathrm{p}(\varphi) - L'(\varphi)\int_0^\infty d\ell \  \ k[\ell-L(\varphi)] P_\mathrm{fast}(\ell|\varphi)\right)P_\varphi \right]~.
\end{equation}
The phase dynamics is thus governed by an effective periodic potential $\mathcal{V}_\mathrm{eff}(\varphi)$ of the form
\begin{align}\label{eq:s_effectiveU}
\mathcal{V}_\mathrm{eff}(\varphi) = -\frac{\Delta G}{2\pi}\varphi + V_\mathrm{p}(\varphi) - \int^\varphi d\varphi' L'(\varphi')\int_0^\infty d\ell \ k[\ell-L(\varphi')] P_\mathrm{fast}(\ell|\varphi').
\end{align}
The potential in Eq.~\eqref{eq:s_effectiveU} can be used to calculate stochastic observables as in the examples shown at Fig.~3, in a similar way as done with the potential in Eq.~\eqref{eq:veff} in the main text. If one takes the limit of $k\to \infty$ finds 
\begin{align}
    - \int^\varphi d\varphi' L'(\varphi')\int_0^\infty d\ell \ \ell^2 k[\ell-L(\varphi')] P(\ell;\varphi') \to \frac12 \kappa \phi_{\cc} L(\varphi)^2 -2 k_\mathrm{B}T\ln L(\varphi),
\end{align}
and thus, we recover Eq.~\eqref{eq:veff}, as expected.

\section{Michaelis-Menten kinetics with crowding as a Markov process}

In this section, we provide a minimal model where the catalytic cycle of enzymes is treated as a thermodynamically consistent Markov process with Michaelis-Menten kinetics and the crowding effects incorporated in the energetics of the system.

An enzymatic cycle can be divided into different stages and intermediate processes, which can be used towards building a Markov process. For simplicity, we consider an enzymatic cycle of three discrete states $\{A,B,C\}$. Each state corresponds to a different chemical state with a corresponding conformational state. In particular; in the state $A$, the enzyme is free to bind a substrate molecule (S), then by transitioning
to $B$ with some probability rate $k_{BA}$, the enzyme forms a complex $ES$. Then, the catalytic substrate-to-product reaction takes place via the $k_{CB}$ rate such that $ES\to EP$, with $EP$ being the $C$ chemical state. Upon releasing $P$, we complete the cycle by going back to the state $A$. Therefore, we have three probabilities $P_A+P_B+P_C=P_{E}+P_{ES}+P_{ES}=1$ and 6 transition rates. For thermodynamic consistency and in the absence of external forces, the various forward and backward rates 
need to satisfy a local detailed-balance condition
\begin{align}
    \frac{k_{\beta \alpha}}{k_{\alpha \beta}} = \exp\left(\beta \Delta \mu_{\beta \alpha}\right),
\end{align}
with $\Delta \mu_{\beta \alpha}$ being the free energy cost for the transition $\alpha \to \beta$. The transition rates for this model can be obtained from the characteristic mass-action law as follows
\begin{align}
    &\Delta \mu_{BA}= k_\mathrm{B}T\ln(k_{BA}[S]/k_{AB}),\\
    &\Delta \mu_{CB}=k_\mathrm{B}T\ln(k_{CB}/k_{BC}),\\
    &\Delta \mu_{AC}=k_\mathrm{B}T\ln(k_{AC}/k_{CA}[P]),
\end{align}
where $[S]$ and $[P]$ are the concentrations of substrate and product molecules, respectively.
To include the effective forces due to the protein condensate, we must include the corresponding work done by the fluctuating forces as derived in Section \ref{sec:coarse-grain}. For the specific $\alpha \to \beta$ transition, this reads 
\begin{align}
W_{\beta \alpha} = \kappa\phi_{\cc} \int_{\ell_\alpha}^{\ell_\beta} d \ell \, \ell = \frac{1}{2}\kappa\phi_{\cc}\left( \ell_\beta^2-\ell_\alpha^2\right),   
\end{align}
where $\ell_{\alpha}$ is the enzyme length in state $\alpha$. We expect to have  $\sum W_{\beta \alpha} = 0$ over a closed cycle.

\begin{figure}[t]
\centering\includegraphics[width=0.55\linewidth]{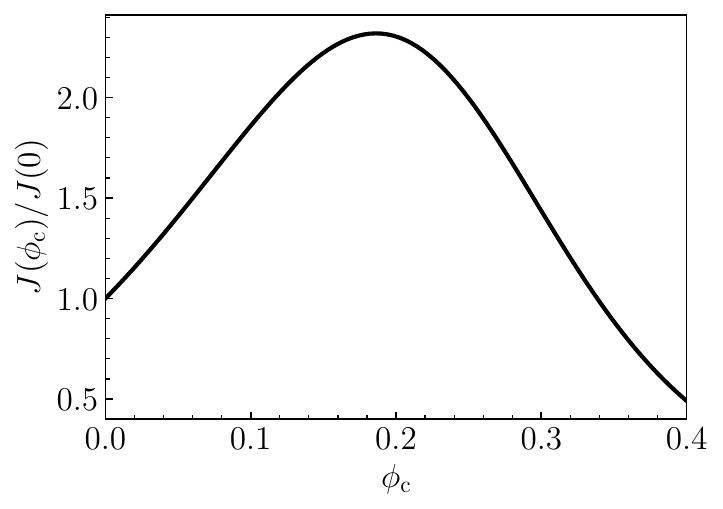}
\caption{The rescaled steady-state current $J(\phi_{\cc})$ of the non-equilibrium enzymatic cycle as a function of the crowder volume fraction $\phi_{\cc}$. The parameters used in the plot are $\theta_{BA}=5/6$, $\theta_{CB}=\theta_{AC} = 1/6$, $[S]=0.1$, $[P]=1$, $\beta \Delta \mu_{BA}=\beta \Delta \mu_{CB} = \beta \Delta \mu_{AC} = 8/3$, $\beta\kappa L_0^2=100 $, $\ell_A= L_0$, $\ell_B=1.1 L_0$, and $\ell_C = 0.95 L_0$.}
\label{fig:supmat_fig}
\end{figure}

As a concrete example, let us assume that $\ell_A= L_0$, $\ell_B=1.1 L_0$ and $\ell_C = 0.95 L_0$. The local detailed-balance conditions are modified as follows
\begin{align}
    &\Delta \mu_{BA}+W_{BA}=k_\mathrm{B}T\ln(k_{BA}[S]/k_{AB}), \\
    &\Delta \mu_{CB}+W_{CB}=k_\mathrm{B}T\ln(k_{CB}/k_{BC}),\\
    &\Delta \mu_{AC}+W_{AC}=k_\mathrm{B}T\ln(k_{AC}/k_{CA}[P]),
\end{align}
and we have the freedom to define the rates as
\begin{align}
    k_{\beta \alpha} = k_{\beta \alpha0}\exp\left(\beta \Delta \mu_{\beta \alpha}\right)\exp\left(\theta_{\beta \alpha} \beta W_{\beta \alpha} \right),
\end{align}
such that $\theta_{\beta \alpha} = 1-\theta_{\alpha \beta}$~\cite{PhysRevResearch.6.L022044}. By solving the master equation in the steady state, we find the current $J$, which represents the cyclic rate, i.e. the number of complete reactions per unit time. Its expression is given as follows
\begin{align}\label{eq:Jdiscr}
    J = \frac{k_{BA} k_{CB} k_{AC}[S]-k_{AB}k_{BC}k_{CA}[P]}{k_{BC}k_{CA}[P]+k_{BC}k_{BA}[S]+k_{AC}k_{BA}[S]+k_{AB}k_{BC}+k_{AB}k_{AC}+k_{CB}k_{AC}+k_{CA}k_{AB}[P]+k_{CA}k_{CB}[P]+k_{BA}k_{CB}[S]}.
\end{align}
As depicted in Fig.~\ref{fig:supmat_fig}, $J(\phi_{\cc})$ has a non-monotonic response as a function of the crowder volume fraction. 

\section{Simulations}

In simulations, the enzyme is modeled as a dumbbell formed by two spheres at positions $\p{R}_1$ and $\p{R}_2$, separated by a length $\ell=|\p{R}_1-\p{R}_2|$. The enzyme also has an internal variable $\varphi$, representing which stage the enzyme is at within its reaction cycle. 
The dynamics of $\varphi$ is determined by $U^\ee(\varphi, \p{R}_1, \p{R}_2)$, with the addition of white noise to represent internal fluctuations of the enzyme within its reaction cycle.
The enzyme subunits try to maintain a separation given by $L(\varphi)$, leading to a force on the subunits given by $F=-k[\ell-L(\varphi)]$.
The enzyme subunits furthermore experience a volume exclusion interaction with the crowder particles (but not with each other). The crowder particles are represented by spheres, and experience a volume exclusion with the enzyme subunits and amongst each other.
For the excluded volume interaction we use a Hertzian form, with the volume exclusion potential between two spheres $i$ and $j$ being given by
\begin{equation}
    V_\text{excl}(Y) = \epsilon (1-Y/\sigma)^{\alpha},
\end{equation}
where $Y = |\p{Y}_i-\p{Y}_j|$, and $\sigma = (\sigma_i+\sigma_j)/2$.
We use $\alpha=2.5$, $\epsilon=36$, and spheres with unit diameters for each enzyme subunit and the crowder particles.

For Fig.~4 in the main text, we use $\beta \Delta G=5$, $v_1=0.5\Delta G$, and $v_2=0.1\Delta G$. For the enzyme dumbbell, we use $k=100$, $L_0=0.5$ and $\lambda=0.2$. 
Each subunit and crowder particle has a diffusion coefficient 0.01, and $\phi$ has a noise strength of 1.
The simulations are performed in two dimensions, to reduce the number of particles required to achieve a high crowding density. Each data point in Fig.~4 corresponds to 100 simulations, simulated for $200,000$ time steps, with a timestep of $\Delta t=0.005$.

\end{document}